\documentstyle[amssymb,preprint,aps]{revtex}

\begin{document}
\title{Anomalous transparency for near-top Bloch states of periodic potential}
\author{A.Zh. Muradyan, G.A. Muradyan}
\address{Department of Physics, Yerevan State University, 1 Alex Manougian,\\
Yerevan 375049 Armenia.}
\maketitle

\begin{abstract}
In the paper are considered stationary (Bloch) states of a particle, in the
field of \ periodic biparabolic type potential. \ It is shown that while the
particle's energy decreases in limits of a single energy band, the
probability of the particle to be in barrier-type region of periodic
potential increases, in contrary to the expected decreasing. \ This
''anomalous'' behavior is more pronounced for the near-top bands and
monotonically decreases for the higher or lower ones.
\end{abstract}

\section{\protect\bigskip Introduction}

Bloch functions are the basis of periodic system theory, particularly of
crystalline solid physics \cite{1} and optical lattices\cite{2}. For a
high-precision approximation of a standing wave (sinusoidal potential), by
one of the authors was suggested a biparabolic form for the potential\cite{3}%
, Bloch functions of which are expressed by confluent hypergeometric
functions. \ The approximation for the potential has been used to calculate
the spontaneous emission in the field of resonant standing wave, the
presence of light-induced anisotropy in that emission was shown \cite{4}.
Also it was used for calculating the temperature of ideal gas BEC in a
periodic field and was shown that the critical temperature decreases with
the deepening of the potential\cite{5}.

In this paper the biparabolic periodic potential is used to investigate the
Bloch states themselves, more concretely, for identifying the behavior of
modulo wave function as a function of energy in the limits of a specific
energy band. \ Specially we are interested in so-called ''underbarrier''
(classically forbidden) states. \ 

The way of dependance of underbarrier states on energy is well-known for a
single potential barrier and for transitions from one energy band to the
neighboring ones: the underbarrier wave function decreases\ conjointly with
the decreasing of energy of a particle. But what is the kind of dependance\
in the limits of a sole (definite) energy band? Surprisingly this question
has slipped from investigators sight; at least in scientific literature
available to us we did not find works concerning this question. \ As a
formal argument for examining this question, for us appears the fact that
Bloch states are generated through the interference of particle de Broglie
waves transmitted and deflected from many (infinite in the limit) potential
barriers, and hence interference, besides the energy, also will play an
essential role in establishment of regularities. Of course, the deflection
and interference of underbarrier states take place in case of a
single-barrier potential too, but they are small there and as a dominating
behavior will appear the energetic behavior. Consequently, the inter-band
regularity\ is not under necessity to be repeated for intra-band states too.
Really, in case of periodic potential, when from different barriers
multiple-wave interference takes place, as a dominating one can appear the
interference behavior ,\ and the dominant role of interference will be
expected for intra-band regularities. Than, if the interference behavior
does not coincide with energy one, the general behavior will mimic the
interference but not energy regularity. \ Our calculations show that indeed,
when particle's energy decreases in limits of a definite band, the modulus
of the wave function of underbarrier (classically forbidden) region does not
decrease, as should be expected from energy reasonings, but increases. The
growth rate is the greatest for the band, nearest to the height of periodic
potential.

\bigskip

\section{Biparabolic potential and near-top Bloch state approximation}

\bigskip Stationary Schr\"{o}dinger equation can be written in a standard
form 
\begin{equation}
\left( \frac{d^{2}}{dz^{2}}+E-V(z)\right) \Psi (z)=0,  \eqnum{1}  \label{1}
\end{equation}
where the particle coordinate $z$ is normalized in units of the potential
period $L$ with coefficient $2\pi $ ($z=2\pi z/L=kz,$ $k$ is the
one-dimensional reciprocal lattice constant) and the full energy $E$ and the
potential energy $V(z)$ are normalized in units of \ ''lattice'' recoil
energy quantum $E_{r}=\hbar ^{2}k^{2}/2M$, where $M$ is mass of the
particle. \ Taking for convenience the zero of the energy on the level of
potential energy minimums, biparabolic potential $V(z)$ takes the form (see
for more details \cite{5} or \cite{3}) 
\begin{equation}
V(z)=\frac{1+(-1)^{m}}{2}V-(-1)^{m}\chi (z-m\pi )^{2},  \eqnum{2}  \label{2}
\end{equation}
where $V$ is the height of the periodic potential, $\chi =2V/\pi ^{2}$, $%
m=0,\pm 1,\pm 2,...,$ and $z$ for each $m$ lays in range $(m-1/2)\pi \leq
z\leq (m+1/2)\pi .$

The form of the potential is given in fig.\ref{Fig.1}, where region $I$ (and
others similar to it) will be named well-type, and region $II$ (and similar
ones) - barrier-type.

Linearly independent solutions of \ Schr\"{o}dinger Eq.(\ref{1}) for
well-type region $I$ are written in form \cite{3},\cite{5} 
\begin{equation}
\varphi _{1}(z)=\exp \left( -\frac{\sqrt{\chi }z_{1}^{2}}{2}\right) \Phi
\left( \alpha ,\frac{1}{2};\sqrt{\chi }z_{1}^{2}\right) ,  \eqnum{3}
\label{3}
\end{equation}
\begin{equation}
\varphi _{2}(z)=z_{1}\exp \left( -\frac{\sqrt{\chi }z_{1}^{2}}{2}\right)
\Phi \left( \alpha +\frac{1}{2},\frac{3}{2};\sqrt{\chi }z_{1}^{2}\right) , 
\eqnum{4}  \label{4}
\end{equation}
where $z_{1}=z-\pi ,$ $\pi /2\leq z\leq 3\pi /2$, $\Phi \left(
..,..;..\right) $ is the confluent hypergeometric function and 
\begin{equation}
\alpha =\frac{1}{4}\left( 1-\frac{E}{\sqrt{\chi }}\right) .  \eqnum{5}
\label{5}
\end{equation}
The total wave function in this region will be 
\begin{equation}
\Psi _{I}(z)=c_{1}\varphi _{1}(z)+c_{2}\varphi _{2}(z),  \eqnum{6}  \label{6}
\end{equation}
where $c_{1}$ and $c_{2\text{ }}$are unknown constant coefficients.

The corresponding solutions for barrier-type region $II$ have the form 
\begin{equation}
\overline{\varphi }_{1}(z)=\exp \left( \frac{i\sqrt{\chi }z_{2}^{2}}{2}%
\right) \Phi \left( \beta ,\frac{1}{2};-i\sqrt{\chi }z_{2}^{2}\right) , 
\eqnum{7}  \label{7}
\end{equation}
\begin{equation}
\overline{\varphi }_{2}(z)=z_{2}\exp \left( \frac{i\sqrt{\chi }z_{2}^{2}}{2}%
\right) \Phi \left( \beta +\frac{1}{2},\frac{3}{2};-i\sqrt{\chi }%
z_{2}^{2}\right) ,  \eqnum{8}  \label{8}
\end{equation}
where $z_{2}=z-2\pi ,$ $3\pi /2\leq z\leq 5\pi /2$ and 
\begin{equation}
\beta =\frac{1}{4}\left( 1-i\frac{E-V}{\sqrt{\chi }}\right) ,  \eqnum{9}
\label{9}
\end{equation}
and the total wave function in this region will be 
\begin{equation}
\Psi _{II}(z)=\overline{c}_{1}\overline{\varphi }_{1}(z)+\overline{c}_{2}%
\overline{\varphi }_{2}(z),  \eqnum{10}  \label{10}
\end{equation}
with yet arbitrary coefficients $\overline{c}_{1}$ and $\overline{c}_{2}.$

With the help of continuity requirements at boundary points $z=3\pi /2$ and $%
5\pi /2$ and Bloch periodicity we get the dispersion relation, which can be
written in form 
\begin{equation}
\cos (2\pi P)=1+2G_{11}(E)G_{22}(E),  \eqnum{11a}  \label{11a}
\end{equation}
or 
\begin{equation}
\cos (2\pi P)=-1+2G_{12}(E)G_{21}(E),  \eqnum{11b}  \label{11b}
\end{equation}
where $P=p/2\hbar k$ is the normalized momentum of particle and 
\begin{equation}
G_{ij}(E)=\left\{ \varphi _{i}(z)\overline{\varphi }_{j}^{/}(z)+\varphi
_{i}^{/}(z)\overline{\varphi }_{j}^{{}}(z)\right\} _{z=\pi /2}\text{ },\text{
\ \ \ \ }i,j=1,2.  \eqnum{12}  \label{12}
\end{equation}
where the prime over the function means derivative with respect to $z.$

The same requirements, as is well known, with additional normalizing
condition 
\begin{equation}
\int_{\pi /2}^{3\pi /2}\left| \Psi _{I}(z)\right| ^{2}dz+\int_{3\pi
/2}^{5\pi /2}\left| \Psi _{II}(z)\right| ^{2}dz=1,  \eqnum{13}  \label{13}
\end{equation}
define the coefficients $c_{1,2}$ and $\overline{c}_{1,2}.$

The edge values of energy bands according to (\ref{11a}) and (\ref{11b}) are
determined as solutions of transcendental equations 
\begin{equation}
G_{ij}(E)=0.  \eqnum{14}  \label{14}
\end{equation}

Analysis of these conditions shows that left- and right-side edge points on
the positive-side axis of quasimomentum (expanded conception of energy
bands) are determined from conditions $G_{11}(E)=0$ and $G_{12}(E)=0$
respectively for $n$ being even, and by conditions $G_{21}(E)=0$ and $%
G_{22}(E)=0$ for $n$ being odd$.$ \ For keeping analogy with the harmonic
potential the band with the smallest energy is designated $n=0.$

For even-numbered bands it is appropriate to express the coefficients by $%
c_{1,}$%
\begin{equation}
c_{2}=\frac{G_{11}(E)}{G_{21}(E)}\frac{e^{i2\pi P}+1}{e^{i2\pi P}-1}c_{1,} 
\eqnum{15}  \label{15}
\end{equation}
\begin{equation}
\overline{c}_{1}=\frac{e^{i2\pi P}+1}{2G_{21}(E)}c_{1,}\text{ \ \ \ \ \ \ \
\ \ \ \ \ \ \ \ \ \ \ \ \ \ \ \ \ \ \ \ \ \ \ \ \ \ \ \ \ \ \ \ \ \ \ \ \ \ }%
\overline{c}_{2}=\frac{e^{i2\pi P}-1}{2G_{22}(E)}c_{1,}  \eqnum{16}
\label{16}
\end{equation}
where $c_{1}$ is determined from the normalizing condition. For odd-numbered
bands it is appropriate to express the coefficients by $c_{2}$; 
\begin{equation}
c_{1}=\frac{G_{21}(E)}{G_{11}(E)}\frac{e^{i2\pi P}-1}{e^{i2\pi P}+1}c_{2,} 
\eqnum{17}  \label{17}
\end{equation}
\begin{equation}
\overline{c}_{1}=\frac{e^{i2\pi P}-1}{2G_{11}(E)}c_{2,}\text{ \ \ \ \ \ \ \
\ \ \ \ \ \ \ \ \ \ \ \ \ \ \ \ \ \ \ \ \ \ \ \ \ \ \ \ \ \ \ \ \ \ \ \ \ \ }%
\overline{c}_{2}=\frac{e^{i2\pi P}+1}{2G_{12}(E)}c_{2,}  \eqnum{18}
\label{18}
\end{equation}
and use the normalizing condition to determine $c_{2.}$

Let's now consider the states, with energies near to the height $V$ of
periodic potential: $E\approx V.$ In fig.\ref{Fig.1} these energies lay
inside the dotted lines. For them the parameter $\beta $ (see (\ref{9})),
which determines the character of Bloch wave functions in barrier-type
region, possesses the value $\beta =1/4.$ \ The corresponding formulas (\ref
{7}) and (\ref{8}) are simplified with help of a well known representation
for Bessel functions $J_{\nu }(x)$ by confluent hypergeometric ones: 
\begin{equation}
J_{\nu }(x)=\frac{1}{\Gamma (\nu +1)}\left( \frac{x}{2}\right) ^{\nu
}e^{-ix}\Phi \left( \frac{1}{2}+\nu ,1+2\nu ,2ix\right) ,  \eqnum{19}
\label{19}
\end{equation}
where $\Gamma (\nu +1)$ is the gamma function. \ To use this formula for (%
\ref{7}) we must choose $\nu =-1/4$, and for (\ref{8}) - $\nu =1/4.$ \ After
corresponding substitutions we arrive to 
\begin{equation}
\overline{\varphi }_{1}(z)=\Gamma \left( \frac{3}{4}\right) \left( \frac{%
\sqrt{\chi }z^{2}}{4}\right) ^{1/4}J_{-1/4}\left( \frac{\sqrt{\chi }z^{2}}{2}%
\right) ,  \eqnum{20}  \label{20}
\end{equation}
\begin{equation}
\overline{\varphi }_{2}(z)=\Gamma \left( \frac{5}{4}\right) z\left( \frac{%
\sqrt{\chi }z^{2}}{4}\right) ^{-1/4}J_{1/4}\left( \frac{\sqrt{\chi }z^{2}}{2}%
\right) .  \eqnum{21}  \label{21}
\end{equation}
To avoid misunderstandings we note, that though linearly independent
solutions $\overline{\varphi }_{1}(z)$ and $\overline{\varphi }_{2}(z)$ for
the considered \ approximation do not depend on energy of the particle, the
total Bloch wave function $\Psi _{II}(z)$ is energy dependant via the
coefficients $\overline{c}_{1}$ and $\overline{c}_{2}$.

As to wave-functions $\varphi _{1}(z)$ and $\varphi _{2}(z)$ of well-type
region, we will use the second Tricomi expansion\cite{6} for them, 
\begin{equation}
e^{-x/2}\Phi (a,\sigma +1;x)=\Gamma (\sigma +1)\left( \varkappa x\right)
^{-\sigma /2}\sum_{n=0}^{\infty }A_{n}(\varkappa ,\lambda )\frac{\sigma +1}{2%
}\left( \frac{x}{4\varkappa }\right) ^{n/2}J_{\sigma +n}(2\sqrt{\varkappa x}%
),  \eqnum{22}  \label{22}
\end{equation}
where $\varkappa =\frac{\left( 1+\sigma \right) }{2}-a,$ $A_{0}(\varkappa
,\lambda )=1,$ $A_{1}(\varkappa ,\lambda )=0,...,$ and will restrict
ourselves with the first member of expansion, which corresponds to
sufficiently deep potentials ($\varkappa \succ 1$). \ The respective Bessel
functions are expressed by elementary trigonometric functions and as a
consequence 
\begin{equation}
\varphi _{1}(z)=\cos (\sqrt{E}z),\text{ \ \ \ \ \ \ \ \ \ \ \ \ \ \ \ \ \ \
\ }\varphi _{2}(z)=\frac{1}{\sqrt{E}}\sin (\sqrt{E}z).  \eqnum{23}
\label{23}
\end{equation}
All the coefficients and dispersion relation $\left( (\ref{11a})\text{ or }(%
\ref{11b})\right) $ also undergo sufficient changes. \ The dispersion
relation, for example, takes the form 
\begin{equation}
\cos (2\pi P)=\frac{\pi /4}{\sin (\pi /4)}\left\{ 
\begin{array}{c}
2u\left[ J_{-1/4}(u)J_{-3/4}(u)-J_{1/4}(u)J_{3/4}(u)\right] \cos (\pi \sqrt{E%
})-J_{-1/4}(u)\times \\ 
J_{1/4}(u)\frac{\pi }{2}\sqrt{E}\sin (\pi \sqrt{E}%
)-4u^{2}J_{-3/4}(u)J_{3/4}(u)\frac{2}{\pi \sqrt{E}}\sin (\pi \sqrt{E})
\end{array}
\right\} ,  \eqnum{24}  \label{24}
\end{equation}
where $u=\pi ^{2}\sqrt{\chi }/8.$

\section{Results of numerical calculations}

To display the sought dependance, that is the behavior of Bloch wave
functions due to changing of the energy in limits of a single (defined)
band, we proceed from the above mentioned near-top approximation. \ With the
help of computer simulations we first determined the depth of potential $V$,
so that the last allowed inner-potential band would be found immediately
near the tops of periodic potential (for fig.(\ref{Fig.2}), with $V=1.4494,$%
the energy of mentioned band lays in limits $E_{\min }=$ $0.3947,$ $E_{\max
}=1.4494,\ $and when $V=18.65$, in limits $E_{\min }=13,64,$ $E_{\max
}=18,65 $). \ Then with the help of formulas (\ref{6}) and (\ref{10}), \
with linearly independent solutions (\ref{7}), (\ref{8}) and (\ref{23}),
coefficients (\ref{15}), (\ref{16}) or (\ref{17}), (\ref{18}), dispersion
relation (\ref{24}) and normalizing condition (\ref{13}), we computed the
sought values of \ $\left| \Psi _{I}(z)\right| ^{2}$ and $\left| \Psi
_{II}(z)\right| ^{2}$ as a function of coordinate $z$ along one spacial
period. \ Repeating these calculations for all the energy values of the
band, we obtain a (continuous) sequence of curves, the juxtaposition of
which along the energy lines gives us the seeking behavior.

In Figs. 2a and 2b we present the results of such calculations for two
different depths of potential; one (Fig.2a) for a\ potential depth,
containing two energy bands, and the other one (Fig.2b) containing four
energy bands.\ So, Fig. 2a illustrates the energy-space distribution of
Bloch states in case\ of relatively shallow potential. Fig. 2b illustrates
the same kind distribution for relatively deep potential. Left half of the
face axis ($z$) corresponds to the well-type region of the potential, and
the right half - to the barrier-type region. \ The energy axis is, as is
seen, directed from the reader (energy grows from face to back).

\ As anomalous one is regarded the behavior of wave function in right-half
side(s) of the figure(s), that is the dynamics of Bloch functions in the
barrier-type regions (see Fig.1) as a function of energy. Obviously is seen
that the moving along the energy axis from the back face to the front face,
which physically means getting deeper into potential, does not decrease but,
quite the contrary, increases the modulo square of the wave function. That
merely means that the probability of the particle to be found deeper under
the potential barrier (staying in the limits of a band) is greater than the
probability to be found at smaller depths! \ This kind of behavior is in
contrary to the regularity for the case of deepening into the periodic
potential over various energy bands, or for the case of a single potential
barrier. $That$ $is$ $why$ $we$ $call$ $the$ $behavior$ $anomalous$.

To avoid possible misunderstanding of the mentioned regularity it will be
noted that there is not any strangeness or abnormality in the fact, that the
wave function in barrier-type (right) space region is essentially greater in
average than in well-type (left) space region; it is simply the echo of
overbarrier reflecting, when velocity of the particle decreases and the wave
function (probability of being) respectively increases, in ''underbarrier''
states.

The next question that is of interest, is how does the rate of anomalous
rising depend on the location of the band relative to the height of periodic
potential. For this kind of calculations the near-top energy approximation,
of course, is not applicable and we had to proceed from the precise
formulas. \ A detailed analysis of the regularities obtained for Bloch state
and comparison with single-barrier case we will give in another publication.
We would like only to note that while moving away from near-top energies as
inside as outside the periodic potential, the rate (scaled in energy units)
of anomalous growth decreases.

Finally we note that we carried out the anomalous behavior also for the
Kronig-Penny's potential, but for that case the anomaly is much weaker, than
for the case of biparabolic potential (see fig.(\ref{Fig.3})). It is easy to
understand since the single barrier transparency in high energy region, and
hence the effectivity of matter-wave interference is bigger for
biparabolic-form potential than for rectangular-form one.

This work was supported by ISTC Grant A - 215 - 99.

\bigskip

\begin{figure}[tbp]
\caption{Biparabolic potential and identification of well-type(I) and
barrier-type(II) regions.}
\label{Fig.1}
\end{figure}

\begin{figure}[tbp]
\caption{Modulo square of Bloch wave functions on the plane energy
(E)-coordinate (z) for a near-top (with $E_{\max }=V$) energy band. The
depth (height\}of potential is $V=1.4494$($a$) and $V=18.65(b),$ scaled in
units of recoil energy $E_{r}=(2\hbar k)^{2}/2M.$ The mentioned feature is
in detail described in the text.}
\label{Fig.2}
\end{figure}

\begin{figure}[tbp]
\caption{Modulo square of Bloch wave functions on the plane energy
(E)-coordinate (z) for a Kronig-Penny potential. The depth of potential is $%
V=1.668$, scaled as in Fig.2$.$}
\label{Fig.3}
\end{figure}

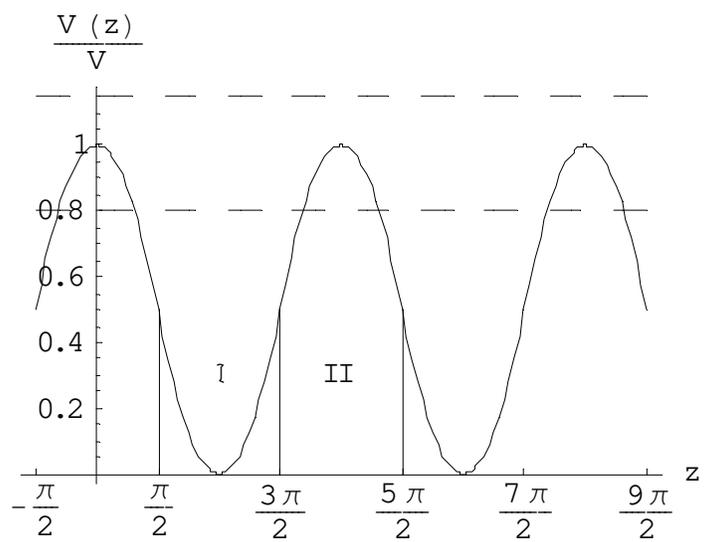
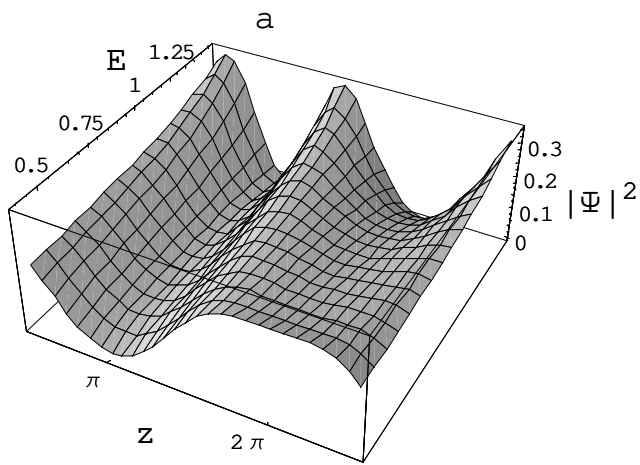
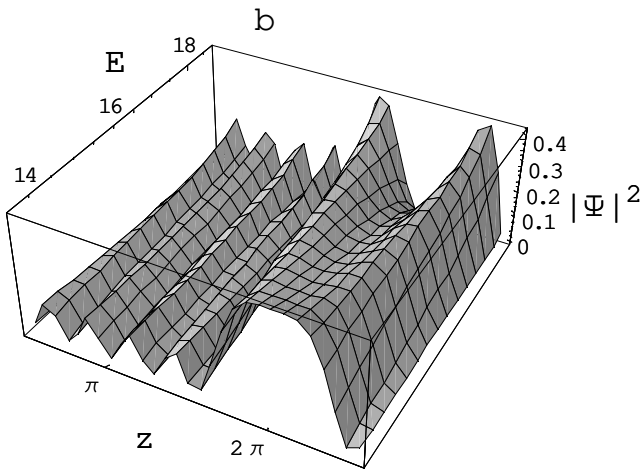
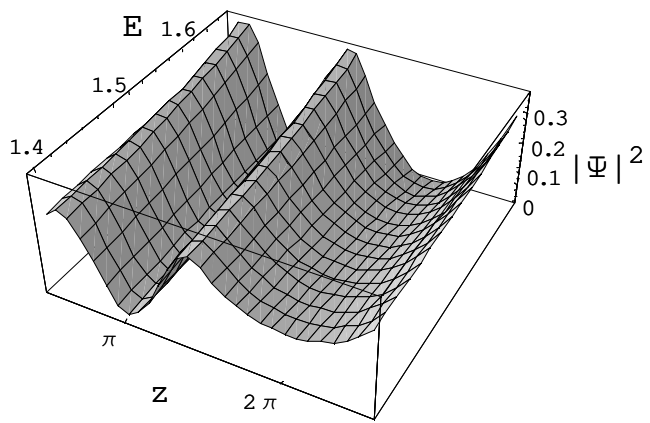
\end{document}